\documentclass{article}
\usepackage{graphicx}
\usepackage{geometry}
\begin{document}

\title{Performance evaluation of the GridFTP within the NorduGrid project}

\author{M.~Ellert$^a$, A.~Konstantinov$^b$, B.~K{\'o}nya$^c$, O.~Smirnova$^c$, A.~W\"{a}\"{a}n\"{a}nen$^d$}
\date{{\footnotesize $^a$Department of Radiation Sciences, Uppsala University,\\
  Box~535, 751 21 Uppsala, Sweden\\
$^b$University of Oslo, Department of Physics,\\
  P.~O.~Box~1048, Blindern, 0316 Oslo, Norway\\
$^c$Particle Physics, Institute of Physics, Lund University,\\
  Box~118, 22100 Lund, Sweden\\
$^d$Niels Bohr Institutet for Astronomi, Fysik og
  Geofysik,\\ Blegdamsvej 17, DK-2100, Copenhagen \O, Denmark}}

\maketitle

\begin{abstract}
  This report presents results of the tests measuring the performance
  of multi-threaded file transfers, using the GridFTP implementation of
  the Globus project over the NorduGrid network resources. Point to
  point WAN tests, carried out between the sites of Copenhagen, Lund,
  Oslo and Uppsala, are described. It was found that multiple threaded
  download via the high performance GridFTP protocol can significantly
  improve file transfer performance, and can serve as a reliable data
  transfer engine for future Data Grids.
\end{abstract}

\section{Introduction}
\label{sec:introduction}

Development of the Grid technologies is the goal of many emerging
projects in distributed, data-intensive computing. The idea of merging
the capacity of computers worldwide is particularly appealing for
tasks, which demand extensive processing of big amounts of data,
located at geographically distributed databases. The Grid provides
innovative solutions, by introducing specific toolkits, which merge
the computers involved into uniform networks.

NorduGrid~\cite{nordugrid} is the project with the aim to create the
Grid computing infrastructure in Nordic countries, making use of the
available middleware. Project participants include universities and
research centers in Denmark, Sweden, Finland and Norway.

A secure, reliable, efficient and high performance data transfer
mechanism over high-bandwidth wide area networks is a key component of
any kind of Grid infrastructure. The Globus metacomputing
project~\cite{globus} has proposed the GridFTP
protocol~\cite{gridftpdef} which contains extensions to the standard
highly popular FTP protocol, in order to meet the requirements of high
performance wide area data movement. Their extended file transfer
protocol supports the following features:
\begin{itemize}
\item Grid Security Infrastructure (GSI)~\cite{gsi}
\item partial file transfer
\item reliable data transfer
\item third-party (from server to server) transfer
\item automatic negotiation of TCP buffer sizes
\item parallel (multi-threaded) data transfer
\item data channel encryption
\end{itemize}

The Globus team has provided software implementation of their protocol
in terms of production libraries and tools. The GridFTP code used was
the alpha-4 release, checked out from the Globus CVS~\cite{gridftprel}.
All the new features except for the TCP buffer negotiation, has been
implemented, and the code became a part of the Globus
Toolkit\texttrademark~2 release~\cite{globus2}.

The purpose of this investigation was to examine the alpha release of
the GridFTP code, test and compare its performance to standard FTP
file transfer within the framework of the NorduGrid project.
Because the GridFTP will be the underlying data transfer engine of
many Grid projects, it is very important to have a first-hand
experience of its performance and capabilities over the NorduGrid
hardware and network resources. Particularly interesting issue is the
expected performance improvements due to the new parallel transfer
mechanism of the GridFTP protocol. It is well-known that the
utilisation of parallel streams in data transfer over high speed WAN
connections is a feasible solution to overcome the TCP buffer size
limitations without the necessity of the modifications of any
sensitive TCP system parameters~\cite{tcp}, therefore parallel streams
can lead to a significant improvement in data throughput. Although
several multi-threaded file transfer clients exist based on the
single-threaded FTP protocol, the Globus Project's GridFTP
implementation is one of a very few, which support parallel streams on
the protocol level. In the tests presented here, the parallel GridFTP
performance over the NorduGrid network have been measured.

\begin{figure}[ht]
\begin{center}
\includegraphics[width=\linewidth]{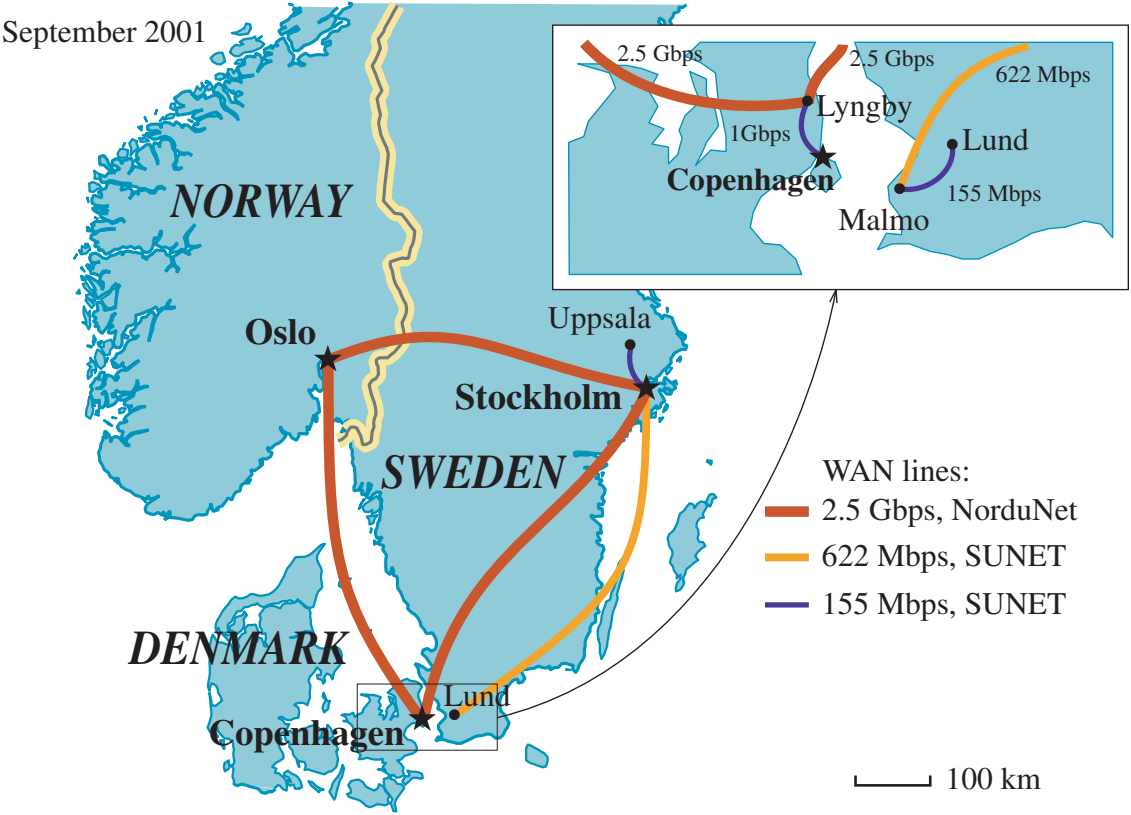}
\caption{Connectivity map of the NorduGrid participating sites. \label{fig:connectivity}}
\end{center}
\end{figure}

\section{Test environment}
\label{sec:environment}
Four NorduGrid sites, Copenhagen, Lund, Oslo and Uppsala, have been
participating in the tests. The sites are connected via the high speed
NORDUnet network~\cite{nordunet}, the actual network configuration is
given in Figure~\ref{fig:connectivity}.

At each site a dedicated Linux server (the Globus gatekeeper
of the local Grid cluster) with a GridFTP server installed was used in the
investigation. The servers of Lund and Uppsala are connected through
a 100~Mbit/s link to the LAN, the Copenhagen server has a
Gigabit connection, while the Oslo server at the time of the
investigation had only a 10~Mbit/s LAN connection.

\renewcommand{\arraystretch}{0.85}
\begin{table}\centering
\caption{Hardware specifications of the GridFTP servers. \label{tab:hardware}}\vspace{0.2cm}
\begin{tabular}{lllll}
\hline\noalign{\smallskip}
GridFTP& CPU & Memory & NIC & LAN  \\
server &  &  &  &  connectivity  \\
\noalign{\smallskip}\hline\noalign{\smallskip}
Lund & P-III 1 GHz & 512 MB & Intel Pro 100 VM & 100 Mbit/s \\
Uppsala & P-III 866 MHz & 512 MB & Intel Pro 100 VE & 100 Mbit/s \\
Copenhagen & 2$\times$P-III 933 MHz & 512 MB & Alteon Ace NIC & 1 Gbit/s \\
 & &  & Gigabit Ethernet & \\
Oslo & P-III 866 MHz & 128 MB & EtherExpress Pro 100 & 10 Mbit/s \\
\noalign{\smallskip}\hline
\end{tabular}
\end{table}

The hardware configurations of the GridFTP servers are listed in
Table~\ref{tab:hardware}; more detailed description can be found on
the NorduGrid Web-site~\cite{resources}.

The Globus Toolkit\texttrademark version 1.1.3b14, shipped with the
{\it globus-url-copy} GridFTP client, was installed and configured on
the machines. Each site ran the GSI-enabled Globus-modified version of
the WU-FTPD server~\cite{wuftpd}. The particular software
configuration of the servers is listed in Table~\ref{tab:software}.

\begin{table}\centering
\caption{Installed software versions.\label{tab:software}}\vspace{0.2cm}
\begin{tabular}{llll}
\hline\noalign{\smallskip}
GridFTP& Platform & Globus  & GridFTP \\
server &  &  Toolkit\texttrademark & software  \\
\noalign{\smallskip}\hline\noalign{\smallskip}
Lund & Mandrake 8.0 & 1.1.3b14 & gsi-wuftpd-0.5 \\
Uppsala & RedHat 7.1 & 1.1.3b14 & gsi-wuftpd-0.5 \\
Copenhagen & RedHat 6.2 & 1.1.3b14 & gsi-wuftpd-0.5 \\
Oslo & RedHat 7.1 & 1.1.3b14 & gsi-wuftpd-0.5 \\
\noalign{\smallskip}\hline
\end{tabular}
\end{table}
\renewcommand{\arraystretch}{1}

\section{Method of measurement}

The {\it globus-url-copy} tool and the {\it gsi-wuftpd} server from
the Globus alpha release 4, were used as the client and server
respectively. A test file of a size of 100~Mbytes was transferred
among the sites, since it was found that this was large enough to
average out most of the network fluctuations. A single test consisted
of repeated number of measurements of the transfer time of the same
file over the same link within a short time interva;. During the
tests, the default TCP settings were not modified, since the purpose
was to study the performance without tuning system parameters.
Throughput measurements with respect to different number of parallel
threads were performed over three different network connections:
\begin{itemize}
\item Lund-Copenhagen (15 router hops)
\item Lund-Uppsala (9 router hops)
\item Lund-Oslo (11 router hops)
\end{itemize}

An example of the command issued to transfer a file in 6 parallel
threads from a local node to Uppsala is:\\
\verb+globus-url-copy -p 6 \ + \\
\hspace*{2cm} \verb+file:/tmp/100mb.tmp + {\tt
gsiftp:}\verb+//grid.tsl.uu.se/tmp/100mb.tmp+ \\
During the investigation, it was found that the background load of the
network link had a significant influence on the download performance:
for example, a single-threaded download time of the test file from
Uppsala to Lund could change from 64s in the morning to 284s in the
afternoon. However, for a reasonable period of time (approximately
half an hour) the network conditions could be considered being stable.
During the measurements, it was found that most of the cases
over any given link the downloading time did not vary more than 5\%,
provided the downloads were made within 30 minutes. Therefore, in
order to eliminate the effect of changing network load, download
times were compared only if the downloads were performed over the
same network link and taken within the same short time interval;
furthermore, all the unreconsctructable and outstanding values were
discarded. For a specific collection of data it was required that the
difference in the download time of the test file measured at the
beginning and at the end of the test session did not exceed 5\%: in
this way one could assume that the test downloads were carried out
over a relatively constant network load of the particular network
link.

\section{Results}
\label{sec:results}

\subsection{FTP clients}
\label{sec:clients}

Before investigating the parallel performance of the GridFTP protocol
and the {\it globus-url-copy} client, tests comparing different Linux
and MS Windows based FTP clients were carried out. Downloads were
performed from the Uppsala server using different FTP clients
installed on a dual booting (Mandrake 8.0 Linux and W2K) P-III 1~GHz
machine in Lund. Under Linux, the FTP, NcFTP, lftp, SSH copy and the
Prozilla~\cite{prozilla} clients were tested. The Prozilla application is a
multi-threaded Linux FTP client. It is capable of opening multiple
connections to a server, where each of the connections downloads a
part of a file, and upon completion of downloads, the partial files
are merged.

\begin{figure}
\begin{center}
\includegraphics[width=0.6\linewidth]{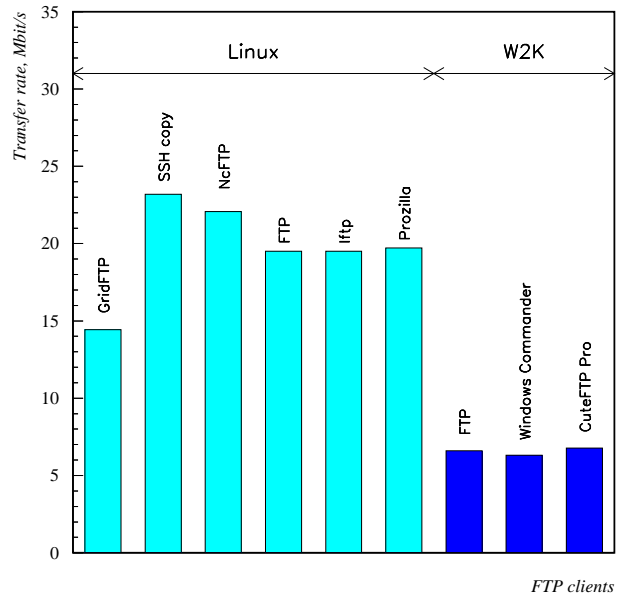}
\caption{Download performance of conventional FTP clients. \label{fig:clients}}
\end{center}
\end{figure}

After rebooting the machine to W2K, the tests were repeated with FTP,
Windows Commander built-in client, and CuteFTP Pro. The latter client
allows up to 4 parallel connections per transferred file.

\begin{figure}
\begin{center}
\includegraphics[width=0.6\linewidth]{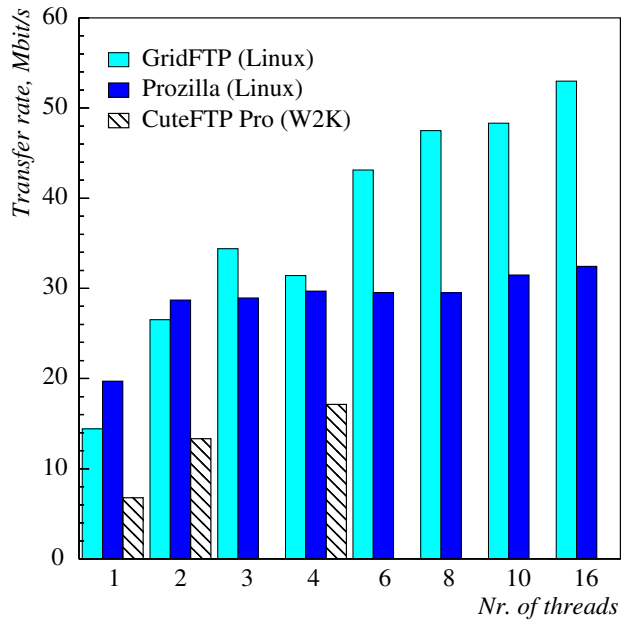}
\caption{Download performance of multi-threaded FTP clients. \label{fig:multiclients}}
\end{center}
\end{figure}

The download performance of the clients were compared to single and
multiple threaded GridFTP downloads. The results are shown in
Figure~\ref{fig:clients} and Figure~\ref{fig:multiclients}. It can be
seen that multi-threaded clients generally improve the transfer time,
while Linux clients routinely outperform those of W2K. The Linux
multi-threaded downloader Prozilla, although providing higher transfer
rates with increasing thread number, generally consumes too much time
during the reconstruction phase, slowing down the performance.
GridFTP provides the fastest downloads already with 4 threads, and can
improve further, providing LAN performance and the network load allow
for it.

\subsection{Lund-Uppsala transfer}
\label{sec:lunduppsala}

The main GridFTP tests were carried out between Lund and Uppsala. At
both sites, the GridFTP servers are connected to their LAN with a fast
100~Mbit/s Ethernet link. Lund has a 155~Mbit/s connection to the
Malm\"{o} backbone, which is connected via a 622~Mbit/s line to
Stockholm, and Stockholm in turn has a 155~Mbit/s line to Uppsala (see
Figure~\ref{fig:connectivity}).

\begin{figure}
\begin{center}
\includegraphics[width=0.7\linewidth]{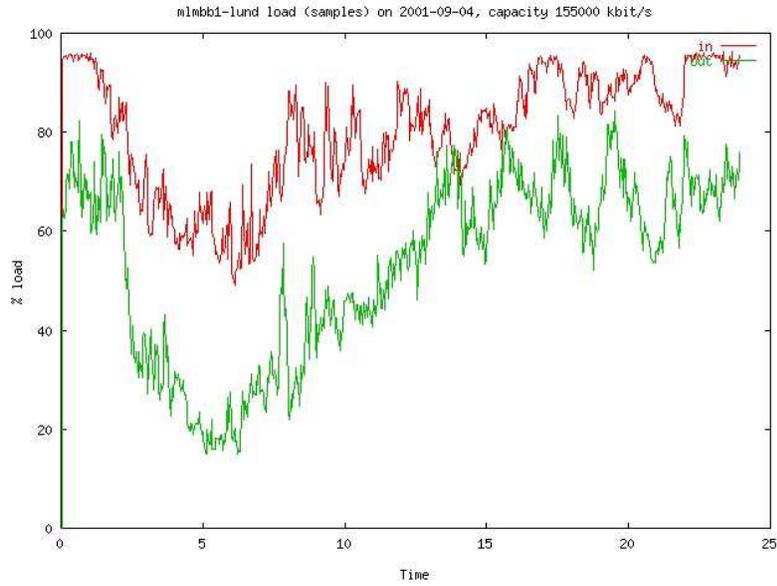}
\caption{Daily network load of the Malmo-Lund link (percentage of total
        capacity, in- and outgoing traffic). \label{fig:malmolund}}
\end{center}
\end{figure}

\begin{figure}
\begin{center}
\includegraphics[width=0.7\linewidth]{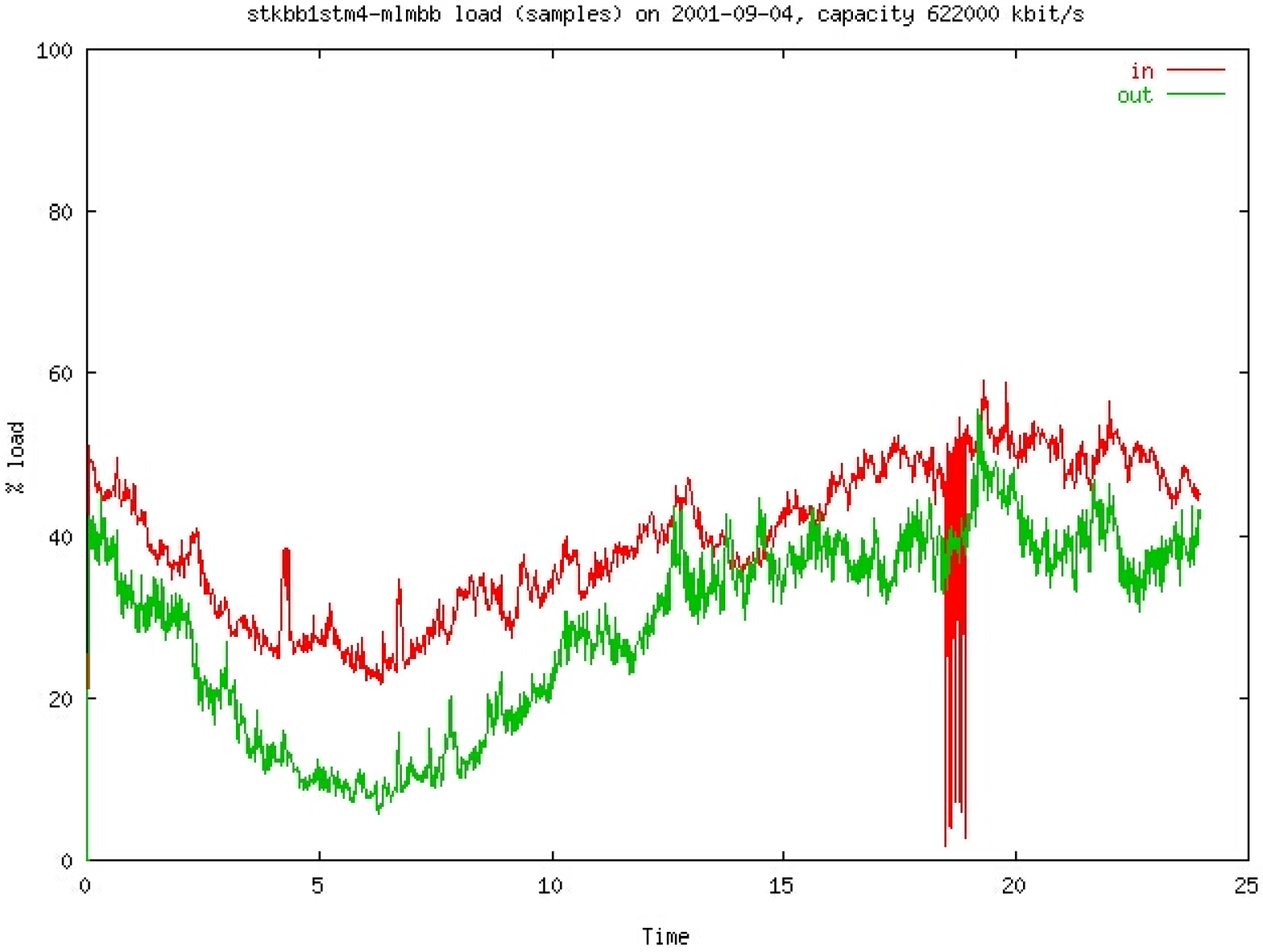}
\caption{Daily network load of the Stockholm-Malmo link (percentage of total
        capacity, in- and outgoing traffic). \label{fig:stkmalmo}}
\end{center}
\end{figure}

\begin{figure}
\begin{center}
\includegraphics[width=0.7\linewidth]{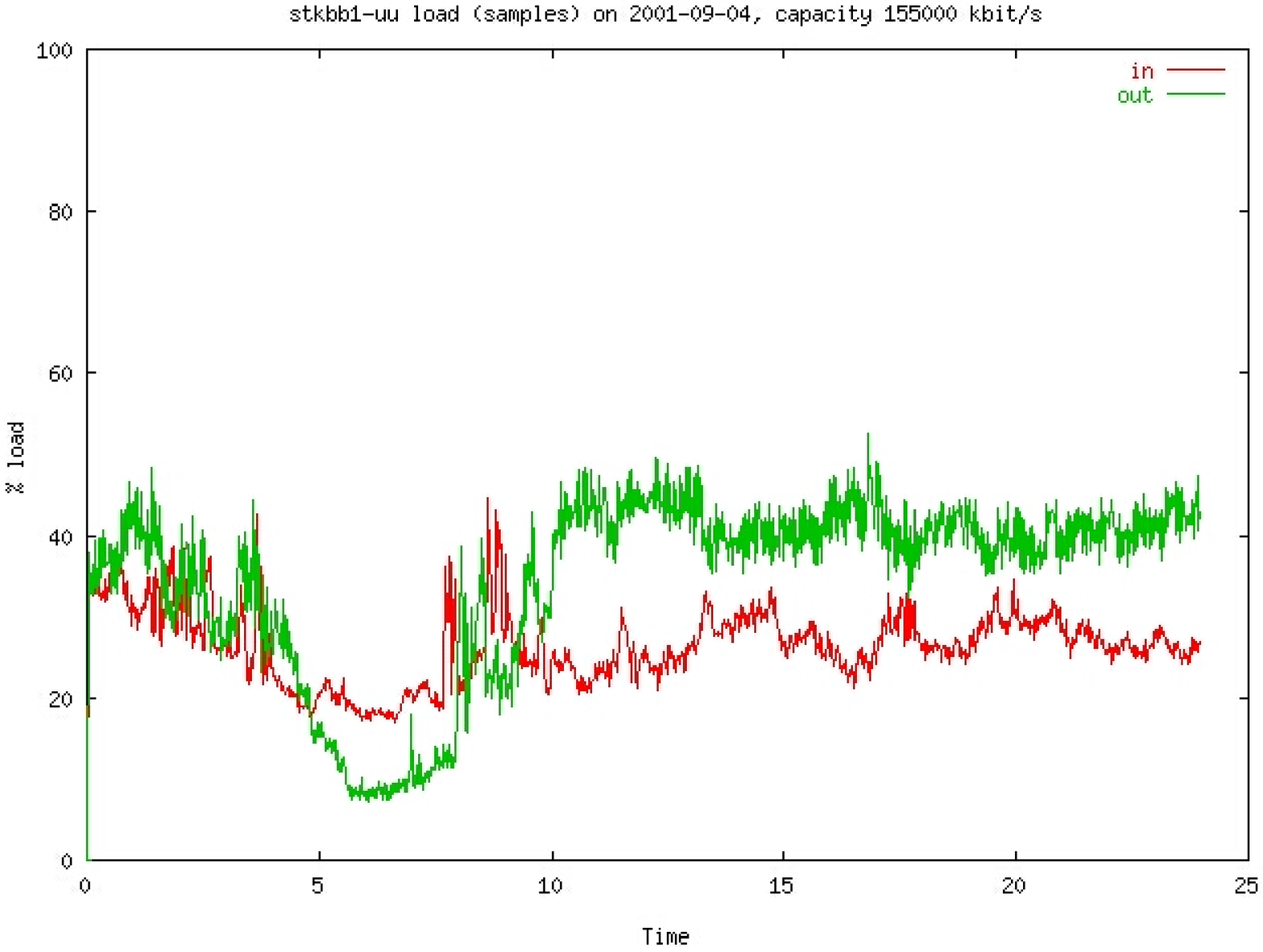}
\caption{Daily network load of the Stockholm-Uppsala link (percentage of total
        capacity, in- and outgoing traffic). \label{fig:stkupps}}
\end{center}
\end{figure}

The network load of the data paths on the day of the measurements is
shown in Figures~\ref{fig:malmolund} to
\ref{fig:stkupps}~\cite{stats}. Most of the day (9:00 thru 1:00 CET),
the Lund-Malm\"{o} link is usually overloaded (running at 90-100\% of
its total capacity), while the other two links have a moderate load of
~40\%.

\begin{figure}
\begin{center}
\includegraphics[width=0.9\linewidth]{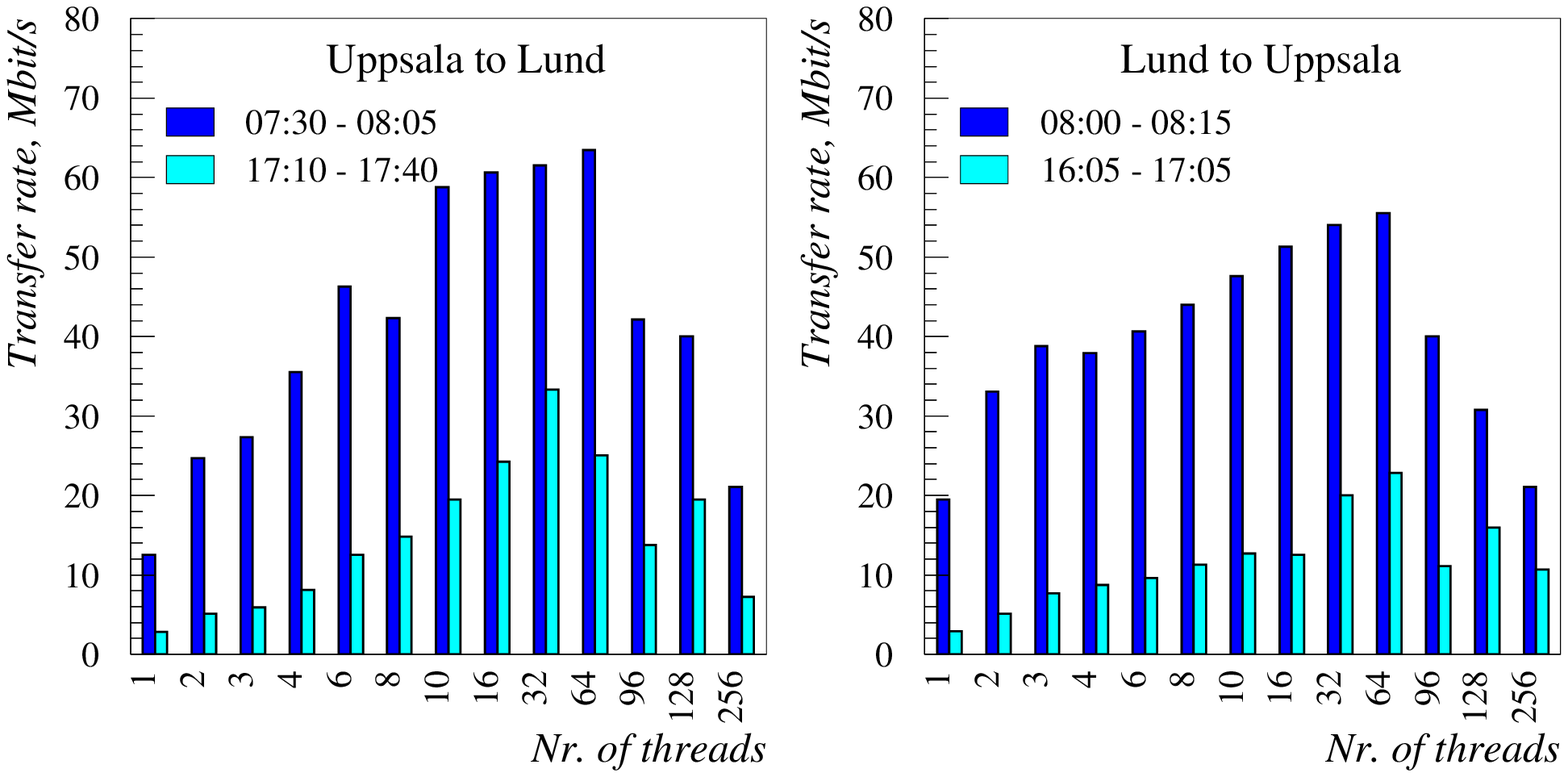}
\caption{Transfer rates for GridFTP downloads between Uppsala and Lund,
        measured at different network load. \label{fig:uu2lu}}
\end{center}
\end{figure}

Two sets of tests were performed: the first series of transfer
measurements were taken in the morning at a relatively low network
load, while the second was performed over a congested network at peak
time. The throughput performance of the GridFTP with respect to the
number of parallel threads via transferring the 100~Mbyte test
file between Lund and Uppsala was measured (see Figure~\ref{fig:uu2lu}).

Usage of parallel threads radically increased the transfer rate both
over the congested and uncongested network, regardless of the load.
The performance was steadily increasing for up to $\sim$64 threads;
over 96 threads instabilities and lower transfer rates were
experienced. Over a congested network, the throughput with 64 threads
increased 780\% and 980\% respectively in the two directions, compared
to the normal single-threaded transfer. In case of the unloaded
network, the transfer rate increased from 20~Mbit/s to 60~Mbit/s (Lund
to Uppsala), which is actually around the maximum throughput of a
100~Mbit/s LAN. This means that over the Lund-Uppsala data path already
the LAN performance represents the bottleneck in the multi-threaded
GridFTP downloads.

\subsection{Lund-Oslo transfer}
\label{sec:lundoslo}

\begin{figure}
\begin{center}
\includegraphics[width=0.6\linewidth]{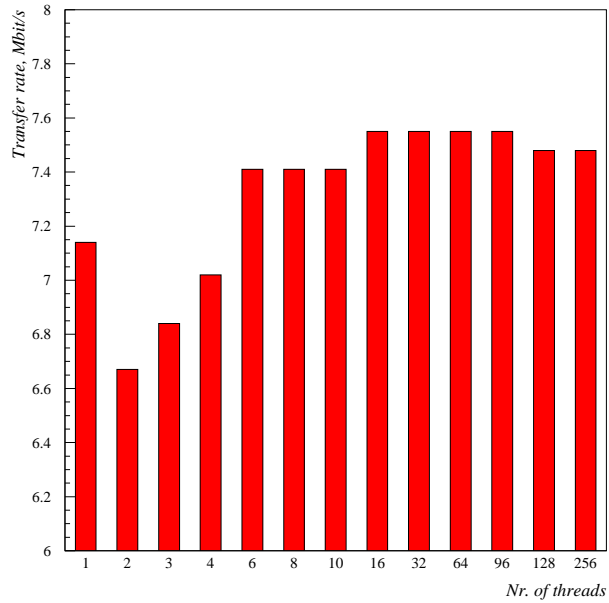}
\caption{GridFTP transfer rates for download from Oslo to Lund. \label{fig:uio2lu}}
\end{center}
\end{figure}

Lund is connected to Oslo through the 155~Mbit/s Lund-Malmo,
2.5~Gbit/s Malmo-Stockholm and 2.5~Gbit/s Stockholm-Oslo links.
However, the 10~Mbit/s LAN connection of the Oslo server was the real
bottleneck on the Lund-Oslo data path.  The result of the multi-threaded
test downloads from Oslo to Lund is shown in Figure~\ref{fig:uio2lu}.
The 7~Mbit/s single-threaded transfer rate is already close to the
10~Mbit/s theoretical maximum, thus the 10~Mbit/s networking
bottleneck seriously limits the functionality of the parallel
downloads.

\subsection{Lund-Copenhagen transfer}
\label{sec:lundcopenhagen}

\begin{figure}
\begin{center}
\includegraphics[width=0.6\linewidth]{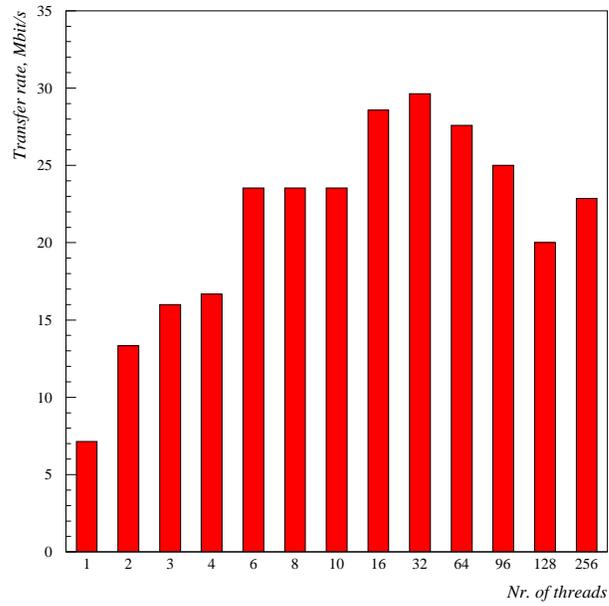}
\caption{GridFTP transfer rates for download from Lund to Copenhagen. \label{fig:lu2nbi}}
\end{center}
\end{figure}

The Lund-Copenhagen datalink, despite the geographical neighbourhood of
the two sites, represents the longest network connection in these
GridFTP tests (15 router hops). The data travel from Lund to
Stockholm, then from Stockholm through a 2.5~Gbit/s link to Lyngby,
the Danish NORDUnet gateway, and finally arrives to Copenhagen (see
Figure~\ref{fig:lu2nbi}). Performing parallel downloads from Lund to
Copenhagen, a throughput gain from 7~Mbit/s to 30~Mbit/s (over a
congested network) was experienced, obtained with 32 threads compared
to the single download rate.

\subsection{Third party transfers}
\label{sec:thirdparty}

\begin{figure}
\begin{center}
\includegraphics[width=0.6\linewidth]{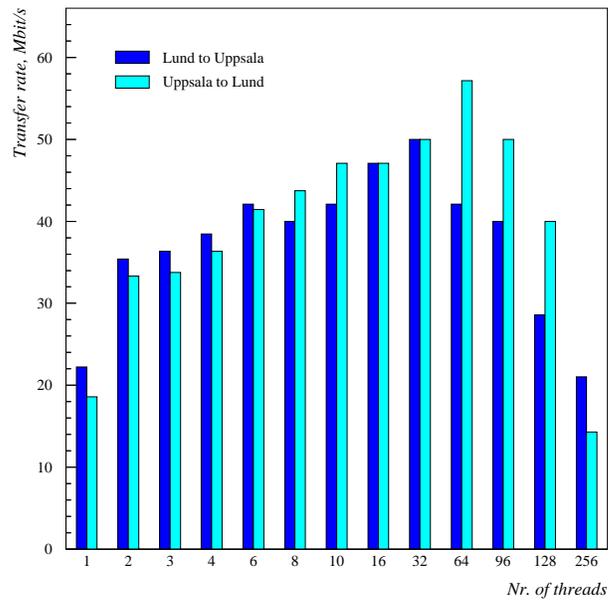}
\caption{GridFTP transfer rates for the third-party (Copenhagen) transfer
        between Lund and Uppsala. \label{fig:uu3lu}}
\end{center}
\end{figure}

Finally, third party transfer tests from Copenhagen (controller)
between the servers of Lund and Uppsala were made (see
Figure~\ref{fig:uu3lu}). A third party transfer implies two FTP
control channels to servers from the controller, and one data channel
between the two servers. The parallel threads resulted in a
performance gain similar to one of the case of the direct transfer
(compare Figure~\ref{fig:uu2lu} and Figure~\ref{fig:uu3lu}). The
transfer rate increased from the single-threaded 15-20 M~bit/s up to
50-55~Mbit/s, achieved with 32 threads.

\subsection{Stability issues}
\label{sec:stability}

In spite of the alpha status of the GridFTP software implementation,
it was found that the {\it globus-url-copy} client and the {\it
  gsi-wuftpd} server are rather stable. However, using more than
hundred parallel streams can sometimes cause instabilities. On several
occasions, with large number of used threads, the start-up phase of
the download froze for several seconds. Restarting the transfer
usually solved the problem. More common unstable behaviour of the
above-100-threads downloads was the extremely slow completion of the
last few hundred kilobytes. On few occasions, the download process in
its finishing state (retrieving the last few hundred bytes) suddenly
restarted from the beginning. A general conclusion is that the
software implementation for above ca. 100 threads became
unreliable. We expect that these instabilities will be resolved with
the final Globus Toolkit\texttrademark~2 release.

\section{Summary}
\label{sec:summary}

In this report the high performance GridFTP data transfer mechanism
was evaluated over the NorduGrid resources. The tests were mainly
focused on the performance gain due to the usage of parallel streams.
Another purpose was to compare the performance of the Globus GridFTP
implementation to several ordinary FTP clients. It was found that the
performance of the conventional FTP clients differ not more than 20
percent and that the single threaded GridFTP could deliver performance
in the same range.

The multi-threaded GridFTP transfers resulted in a remarkable
performance increase of about 600-800 percent, compared to a single
threaded GridFTP (or conventional FTP) downloads. Parallel threads
lead to increased throughput over both congested and unloaded
networks. It's worth to point out that in some of the cases the LAN or
the actual hardware configuration became the bottleneck in the GridFTP
tests. Parallel third party (server to server) transfers were tested
and similar performance enhancement to direct transfers was achieved.
It is clearly unfair towards other users to execute a multi-threaded
transfer over a congested network, as it consumes bandwith
proportional to the number of threads. However, if a big bandwith is
available, like in the case of a fat long dedicated line, a
multi-threaded transfer is the only way to use all the available
capacity without changing the system parameters.

The tests were by no means exhaustive, many other cases could have
been considered. The NorduGrid network resources are going to be
considerably upgraded (the Lund-Malm\"{o} and the Copenhagen-Lyngby
links) and the tests are planned to be extended using the upgraded
network; moreover a dedicated 1GBit/s CERN-Copenhagen link is about to
be set up for tests purposes in the near future.

The alpha status software is proved to be relatively stable, since
irregularities occurred only for downloads involving exaggerated
amount of threads (above 100). After these tests were done, the Globus
Toolkit\texttrademark~2 was released, which includes an upgraded
version of GridFTP. The conclusion is that the multi-threaded GridFTP
significantly boosted the data throughput over the NorduGrid network
and it is certainly a very promising solution for high performance
data transfer.

\section{Acknowledgements}
The NorduGrid project is supported by the Nordunet2 programme,
financed by the Nordic Council of Ministers and by the Nordic
Governments.



\end{document}